\documentclass[prb,twocolumn,amsmath,amssymb,floatfix]{revtex4}
\usepackage{graphicx}

\newcommand{\be}{\begin{eqnarray}}
\newcommand{\ee}{\end{eqnarray}}
\newcommand{\bfr}{{\bf r}}

\newcommand{\bfp}{{\bf p}}
\newcommand{\bfk}{{\bf k}}
\newcommand{\bfE}{{\bf E}}
\newcommand{\bfd}{{\bf d}}
\newcommand{\bfe}{{\bf e}}
\newcommand{\bfJ}{{\bf J}}

\newcommand{\tlV}{\tilde{V}}

\newcommand{\tlx}{\tilde{x}}
\newcommand{\tly}{\tilde{y}}
\newcommand{\tlz}{\tilde{z}}

\newcommand{\tlbfr}{\tilde{\bf r}}
\newcommand{\tlbfk}{\tilde{\bf k}}

\newcommand{\wbe}{\begin{widetext}}
\newcommand{\wee}{\end{widetext}}
\newcommand{\oncite}{\onlinecite}

\begin{document}

\title{Interaction induced ferro-electricity in the rotational states of polar molecules}

\author{Chien-Hung Lin$^1$, Yi-Ting Hsu$^1$, Hao Lee$^{1}$, and Daw-Wei Wang$^{1,2}$}

\affiliation{
$^{1}$ Physics Department, National Tsing-Hua University, Hsinchu 300, 
Taiwan
\\
$^{2}$ Physics Division, National Center for Theoretical Sciences,
Hsinchu 300, Taiwan
}

\date{\today}

\begin{abstract}
We show that a ferro-electric quantum phase transition can be driven 
by the dipolar interaction of polar molecules in the presence a micro-wave field. The obtained ferro-electricity crucially depends on the harmonic confinement potential, and the resulting dipole moment persists even when the external field is turned off adiabatically. The transition is shown to be second order for fermions and for bosons of a smaller permanent dipole moment, but is first order for bosons of a larger moment. Our results suggest the possibility of manipulating the microscopic rotational state of polar molecules by tuning the trap's aspect ratio (and other mesoscopic parameters), even though the later's energy scale is smaller than the former's by six orders of magnitude.
\end{abstract}

\maketitle
The successful realization of high phase-space-density polar molecules
in the JILA group [\oncite{JILA}] opens a new direction of strongly 
correlated quantum gases: ultracold polar molecules.
Many experimental [\oncite{others}] and theoretical works [\oncite{demille,zoller,wang_dipolar_liquid_Santos,ferro_Melo}] have been carried out in recent years. Most many-body physics proposed so far are based on the situation when the inter-molecule effective interaction is predetermined (by a separated calculation) [\oncite{eugene}] and cannot be affected by the many-body physics. On the other hand, the reverse operation, i.e. manipulating the molecular rotational state by changing a mesoscopic parameter (say trapping frequency or particle number etc.), is still unfeasible, because the rotational energy is about six orders of magnitude larger than the interaction energy in a dilute gas. This makes systems of polar molecules very different from spinor atoms [\oncite{spinor}], although they both have rich internal states and long-ranged dipolar interaction. For the same reason, a spontaneous long-ranged order of molecular rotational state seems impossible either [\oncite{note_ferro}], while the related many-body phenomena, like (anti-)ferro-magnetism and multi-ferroics [\oncite{multiferro}], have been important subjects of solid state physics for decades.
\begin{figure}
\includegraphics[width=8cm]{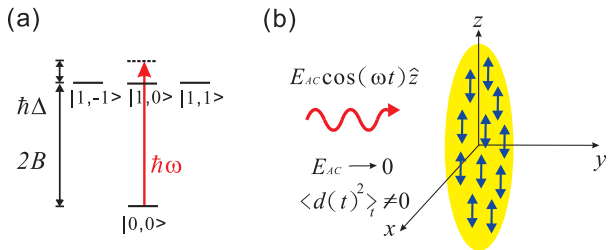}
\caption{(Color online) (a) The lowest two rotational eigenstates of a polar molecule. Here we consider a linearly polarized AC field to couple state ($|g\rangle=|0,0\rangle$) and state ($|e\rangle=|1,0\rangle$) with detuning $\Delta$. (b) Schematic figure for the ferro-electricity in the presence of a microwave field. The blue up-down arrows indicate the oscillating dipole moments.
}
\label{system}
\end{figure}

In this paper, we demonstrate that the rotational state of polar molecules can be manipulated by mesoscopic parameters through an external AC field [\oncite{demille}]: When the trapping potential is elongated and the field frequency is tuned close to the rotational energy (see Fig. \ref{system}(a)), the internal rotational state of molecules can all be "locked" to each other by interaction, leading to a spontaneous mesoscopically ordered state, ferro-electricity. Different from the standard dressed state in a single molecule picture [\oncite{ferro_Melo}], the interaction-induced macroscopic dipole moment here sustains even when the external field is turned off adiabatically. We calculate the quantum phase diagrams for both bosonic and fermionic polar molecules in an elongated ellipsoid trap. We then discuss the nature of the phase transition and how it can be tuned via mesoscopic parameters, like trap's aspect ratio or the detuning of an external field.

The system Hamiltonian of $N$ polar molecules includes both the rotational and the orbital degrees of freedom:
$H_{\rm tot}=\sum_i^N H_{{\rm rot},i}+H_{\rm orb}$. Here $H_{{\rm rot},i}=B\bfJ_i^{2}- \bfd_i\cdot \bfE(t)$ describes the rotational state of the $i$th molecule with ${\bf J}_i$ and $B$ being the angular momentum operator and the rotational constant respectively. $\bfE(t)$ is the external field, and $\bfd_i$ is the electric dipole moment. The orbital part of the Hamiltonian is given by
$H_{\rm orb}\equiv\sum_{i=1}^N\left[\frac{\bfp_i^2}{2m}+V_{\rm t}(\bfr_i)\right]
+\frac{1}{2}\sum_{i\neq j}^N\left[\frac{\bfd_{i}\cdot \bfd_{j}
-3(\bfd_i\cdot\bfe_{ij})(\bfd_{j}\cdot\bfe_{ij})}
{|\bfr_i-\bfr_j|^{3}}\right]$,
%
%
%
where $\bfr_i$ and $\bfp_i$ are the position and momentum operators, and $m$ is the molecule mass. $N$ is the total number of particles and $V_{\rm t}(\bfr)=\frac{1}{2}m(\omega_\rho^2(x^2+y^2)+\omega_z^2 z^2)$ is the external trapping potential. The second term is the dipolar interaction with $\bfe_{ij}\equiv \bfr_{ij}/|\bfr_{ij}|$ being the unit vector of the relative position, $\bfr_{ij}\equiv\bfr_i-\bfr_j$. Following the spirit of the mean-field approximation, without losing generality, we can approximate the ground state wavefunction by a factorizable form:
$\Psi(1,2,\cdots,N)=\Phi_{\rm orb}(\theta;\bfr_1\cdots\bfr_N)\otimes\prod_{i=1}^N
|\theta_i\rangle_i$,
where $\Phi_{\rm orb}(\theta;\bfr_1\cdots\bfr_N)$ is the orbital wavefunction with a variational parameter, $\theta_i$, and $|\theta_i\rangle_{i}$ describes the rotational state of the $i$th particle as more well-defined below. 

In this paper, we are considering the situation when a linearly polarized AC field is applied, $\bfE(t)=E_{\rm AC}\cos(\omega t)\hat{z}$ with an angular frequency $\omega$, to couple rotational states $|g\rangle=|0,0\rangle$ and $|e\rangle=|1,0\rangle$ (see Fig. \ref{system}(a)). 
The single molecule state can be solved by the Floquet theory [\oncite{floquet}] within the rotating wave approximation, and the following time-independent Hamiltonian (in the truncated rotating basis, $\{|g\rangle,e^{-i\omega t}|e\rangle\}$) can be obtained:
\be
\tilde{H}_{{\rm rot},i}
&=&-\frac{\hbar\Delta}{2}-\frac{\hbar\Delta_\Omega}{2}
\left[\begin{array}{cc}
\cos\alpha & \sin\alpha \\
\sin\alpha & -\cos\alpha 
\end{array}\right],
\label{H_rot}
\ee
where $\Omega\equiv d_0E_{\rm AC}/\hbar$ is the Rabi frequency,  
$\Delta\equiv \omega-2B/\hbar$ is the detuning, $\Delta_\Omega\equiv\sqrt{\Delta^2+\Omega^2}$, and $\tan\alpha\equiv\Omega/\Delta$. 
Here $d_0\equiv |\langle g|\bfd_i|e\rangle|$ is the permanent dipole moment. According to the Floquet theorem, a general rotational state can be expressed to be 
$|\theta_i\rangle_i=\cos\left(\frac{\theta_i}{2}\right) |v_+\rangle\,e^{-iE_+t/\hbar}
+\sin\left(\frac{\theta_i}{2}\right)|v_-\rangle\,e^{-iE_-t/\hbar}$,
where $E_\pm=\frac{-\hbar}{2}\left(\Delta\pm\sqrt{\Delta^2+\Omega^2}\right)$ are the eigenenergies, and $|v_+\rangle=-\sin(\alpha/2)|g\rangle+\cos(\alpha/2)
e^{-i\omega t}|e\rangle$ and $|v_-\rangle=\cos(\alpha/2)|g\rangle+\sin(\alpha/2)
e^{-i\omega t}|e\rangle$ are the corresponding eingenstates of $H_{{\rm rot},i}$. After a straightforward calculation, we can obtain the rotational energy per particle, $E_{\rm rot}(\theta)/N=-\frac{\hbar\Delta}{2}-\frac{\hbar\Delta_{\Omega}}{2}
\sum_{i=1}^N\cos\theta_i$, and the averaged dipole moment (in the limit of small $\Delta_\Omega/\omega$):
$\langle\bfd_i(t)\rangle\equiv\langle{}_i\theta_i|\bfd_i|\theta_i\rangle_i=d_0\sin(\theta_i-\alpha)\cos(\omega t)\hat{z}$,
which is oscillating in time. If $\hbar\Delta_\Omega$ is much larger than any other orbital energy scale, the rotational energy is then always minimized at a single dressed-state, $|v_+\rangle$ (i.e. $\theta_i=0$) with a finite rotating dipole moment. However, this state alone is not a true ferro-electric state [\oncite{note_ferro}], no matter what kind of orbital wavefunction it has, because the dipole moment disappears in the limit of zero external field (i.e. $\alpha\to 0$). However, the situation can be totally different when $\Delta$ is tuned to be comparable to the interactional energy, as shown in more details below.

Before further investigation, we note that the $\theta_i$, in principle, can depend on the confinement geometry: when the molecule cloud is elongated along the $z$ axis (i.e. the polarization direction), $\theta_i$ will tend to be uniform in space in order to gain the interaction energy. However, if the cloud is oblate in the $x-y$ plane, $\theta_i$ can become non-uniform in space, i.e. forming a domain wall structure or a rotation-orbital texture [\oncite{texture}]. Since here we concentrate on the ferro-electric order rather than the complete phase diagram, we will always consider a highly elongated trap and assume a uniform polarization for simplicity. A full theory including non-uniform polarization will be presented in the future.

\begin{figure}
\includegraphics[width=8.5cm]{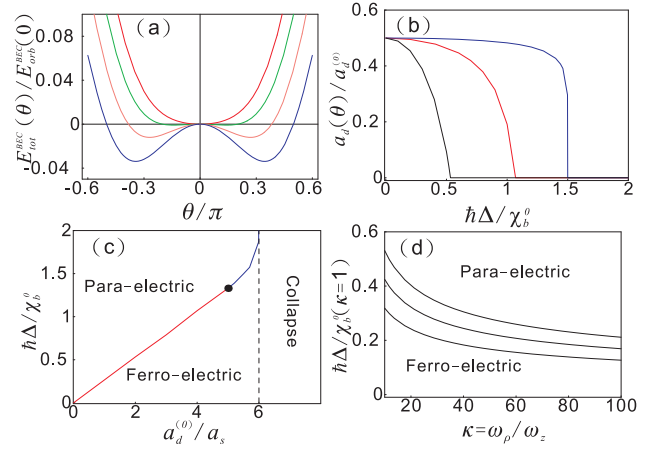}
\caption{Results of bosonic polar molecules: (a) Total groud state energy as a function of $\theta$. Curves from top to bottom are respectively for $\hbar\Delta/\chi_b^0=6$, 5, 4, and 3. (b) The calculated effective dipolar length, $a_d(\theta)$, as a function of the detuning, $\Delta$. Curves are for $a_d^{(0)}/a_s=5.5$, 4, and 2, respectively from top to bottom. (c) Quantum phase diagram as a function of $a_d^{(0)}$ and $\Delta$. The para- to ferro-electricity transition is second order for $a_d^{(0)}/a_s<5$, while it is first order for $a_d^{(0)}/a_s>5$. When $a_d^{(0)}/a_s>6$, the condensate becomes unstable toward collapse. (d) Critical detuning for the second order transition line as a function of trapping aspect ratio, $\kappa$. Curves from top to bottom are for $a_d^{(0)}/a_s=5$, 4, and 3. Note that we have taken $\Omega\to 0$ for all results shown here.
}
\label{phase_boson}
\end{figure}
Within such approximation, the interaction strength can be characterized by the following dipolar length scale in the dilute limit [\oncite{zoller}],
\be
a_d(\theta) &\equiv & \frac{m\langle\langle\bfd(t)\rangle\cdot\langle\bfd(t)\rangle\rangle_t}{\hbar^2}=\frac{a_d^{(0)}}{2}\sin^2(\theta-\alpha),
\label{a_d}
\ee
where $a_d^{(0)}\equiv m d_0^2/\hbar^2$, and 
$\langle\cdots\rangle_t\equiv\frac{1}{\tau}\int_0^\tau dt\cdots$ is the time average over many periods, $\tau\gg 2\pi/\omega$. To study its effect on the orbital many-body wavefunction, we first consider bosonic molecules at zero temperature, where the orbital condensate wavefunction can be well-approximated by a Gaussian type form: $\Psi(\bfr)=\frac{N^{1/2}}{\pi^{1/4}R_\rho R_z^{1/2}}e^{-(\rho^2+z^2/\beta^2)/R_\rho^2}$ with $R_\rho$ being the horizontal radius and $\beta$ being the aspect ratio. The orbital condensate energy, $E^{\rm BEC}_{\rm orb}(\theta)$, then becomes [\oncite{3d_dipolar_wang,Cr}]
\be
\frac{E_{\rm orb}^{\rm BEC}(\theta)}{N\hbar\omega_\rho/2}&=&
\frac{a_\rho^2}{R_\rho^2}\left(1+\frac{1}{2\beta^2}\right)+\frac{R_\rho^2}{a_\rho^2}
\left(1+\frac{\beta^2}{2\kappa^2}\right)
\nonumber\\
&&+\frac{2 N a_s a_\rho^2}{\sqrt{2\pi}\beta R_\rho^3}
+\frac{16A_2(\beta)}{3\sqrt{10\pi}}
\frac{N a_d(\theta)a_\rho^2}{R_\rho^3},
\label{E_BEC}
\ee
where $a_\rho=\sqrt{\hbar/m\omega_\rho}$ is the oscillator length, and $a_s$ is the $s$-wave scattering length. $\kappa\equiv\omega_\rho/\omega_z$.  

Now we minimize the total energy, $E^{\rm BEC}_{\rm tot}(\theta)=E_{\rm rot}(\theta)+E^{\rm BEC}_{\rm orb}(\theta)$, with respect to variational parameters, $R_\rho$, $\beta$, and $\theta$. For an elongated trap (i.e. $\beta\gg 1$), we can use $A_2(\beta)\to\frac{-\sqrt{5}}{8\beta}$ and therefore the interaction energy term of Eq. (\ref{E_BEC}) becomes the same as the result of a non-dipolar condensate with an effective length, $a_{\rm eff}(\theta)\equiv a_s-a_d(\theta)/3$. Within the Thomas-Fermi limit (i.e. $N\gg 1$ and hence the kinetic energy is negligible), the minimum energy at a given $\theta$ can be done analytically [\oncite{BEC_book}]. We obtain $\beta=\kappa\equiv\omega_\rho/\omega_z$, and 
\be
\frac{E^{\rm BEC}_{\rm orb}(\theta)}{N}
=\frac{5}{4}\left(\frac{2}{\pi}\right)^{1/5}\hbar\bar{\omega} 
\left(\frac{N a_{\rm eff}(\theta)}{\bar{a}}\right)^{2/5},
\label{E_BEC_orb}
\ee
where $\bar{\omega}\equiv (\omega_\rho^2\omega_z)^{1/3}$ and $\bar{a}\equiv\sqrt{\hbar/m\bar{\omega}}$. It is easy to see that the system becomes unstable toward collapse when $a_{\rm eff}(\theta)<0$, giving the same condition derived in magnetic atoms [\oncite{Cr}]. Taking zero field limit and expanding $E^{\rm BEC}_{\rm tot}(\theta)$ to the leading order of small $\theta$, we obtain
\be
\frac{E^{\rm BEC}_{\rm tot}(\theta)}{N}
&\approx&\left(\frac{\hbar|\Delta|}{4}-\chi_b\right)\theta^2
+\left(\lambda\chi_b-\frac{\hbar|\Delta|}{48}\right)\theta^4,
\label{E_tot_b}
\ee
after shifting the total energy to zero at $\theta=0$. Here we define $\chi_b\equiv\frac{a_d^{(0)}}{15a_s}\chi_b^0$ to be the "electric susceptibility" and $\lambda\equiv \frac{1}{3}-\frac{a_d^{(0)}}{20 a_s}$. Here $\chi_b^0\equiv E^{\rm BEC}_{\rm orb}(0)N^{-1}$. It is easy to see that a second order quantum phase transition from a para-electricity toward a ferro-electricity can take place when $|\Delta|<\Delta_c\equiv 4\chi_b/\hbar$ and when $\lambda\chi_b-\frac{\hbar\Delta_c}{48}>0$ (i.e. $a_d^{(0)}<5a_s$). In Fig. \ref{phase_boson}(a), we show $E^{\rm BEC}_{\rm tot}(\theta)$ as a function of the variational parameter, $\theta$, near the phase transition of $a_d^{(0)}/a_s=2$. However, when $a_d^{(0)}>5 a_s$, the coefficient of the $\theta^4$ term becomes negative, leading to a first order phase transition instead, see Fig. \ref{phase_boson}(b). In Fig. \ref{phase_boson}(c), we show the calculated quantum phase diagram for such elongated condensate, as a function of $a_d^{(0)}$ and $\Delta$. A critical point is at $a_d^{(0)}/a_s=5$ and $\hbar\Delta/\chi_b^0\sim 1.33$. When $a_d^{(0)}/a_s>6$, the condensate becomes collapse due to the strong attractive dipolar interaction. Note that since $E^{\rm BEC}_{\rm orb}(0)=N\chi_b^0$ depends on the trap's aspect ratio, $\kappa=\omega_\rho/\omega_z$, we can also derive the critical detuning as a function of $\kappa$ to be: $\hbar\Delta_c= \frac{4a_d^{(0)}}{15a_s}\chi_b^0=\frac{4a_d^{(0)}}{15a_s}\chi_b^0(\kappa=1)\kappa^{-2/5}$, and therefore the ferro-electricity can be also tuned easily by changing the trap's aspect ratio as shown in Fig. \ref{phase_boson}(d). Note that the decay of $\Delta_c$ for larger $\kappa$ is simply due to the reduced particle density. Taking $^{39}$K$^{87}$Rb as an example, we have $d_0=0.5$ Debye and hence $a_d^{(0)}=0.47$ $\mu$m. Using $a_s=0.2$ $\mu$m, we can easily obtain $\Delta_c=2\pi\times 250$ Hz for $N=10^4$, $\omega_\rho=2\pi\times 100$ Hz, and $\kappa=100$.  

\begin{figure}
\includegraphics[width=8cm]{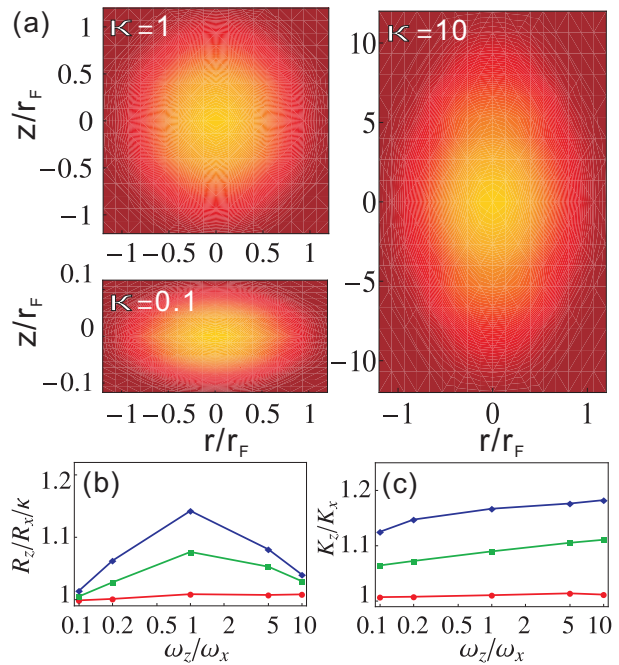}
\caption{(a) Zero temperature density profile (after column integration)
for fermionic polar molecules, $^{40}$K$^{87}$Rb, in three different trap's aspect ratio, $\kappa$s. Here we use $N=3\times 10^5$, $|\langle\bfd\rangle|=0.5$ Debye, and a fixed $\omega_\rho=2\pi\times 100$ Hz. (b) and (c) are renormalized cloud's aspect ratio in the spatial and momentum density distribution as a function $\kappa$. $|\langle\bfd\rangle|=0.1$, 0.3 and 0.4 Debye from bottom to top. 
}
\label{fermi_gas}
\end{figure}
Now we turn to systems of fermionic molecules. To the best of our knowledge, so far only a variational study [\oncite{HanPu}] and Fermi-liquid theory in a uniform system [\oncite{Bohn}] are studied in the literature, while a full self-consistent Hartree-Fock theory with the local density approximation (SCHF+LDA) have not been investigated in the literature. It is therefore worthy to show our results on this aspect first. Within such HF+LDA approximation, the orbital energy can be calculated to be
\be
E_{\rm orb}^{\rm FL}(\theta)&=&\int_{\bfr,\bfk}
\left[\epsilon_\bfk^0+V_{\rm t}(\bfr)+\frac{1}{2}
\Sigma(\bfk,\bfr)\right]f_\bfk(\bfr),
\label{E_FL}
\ee
where $\Sigma(\bfk,\bfr)=\int_{\bfr',\bfk'}
\left[V_d(\bfr-\bfr')-\tlV_d(\bfk-\bfk')\delta(\bfr-\bfr')\right]
\times f_{\bfk'}(\bfr')$is the static local self-energy, and $f_\bfk(\bfr)\equiv
\left[\exp\left((\epsilon_\bfk^0
+V_{\rm t}(\bfr)+\Sigma(\bfk,\bfr)-\mu)/k_BT\right)+1\right]^{-1}$ is 
Fermi distribution at temperature $T$. $\int_{\bfr,\bfk}=\int\int\frac{d\bfr d\bfk}{(2\pi)^3}$ is the integration over the whole phase space.
$V_d(\bfr)=|\langle\bfd\rangle|^2(r^2-3z^2)|\bfr|^{-5}$ is dipolar interaction and $\tlV_d(\bfk)$ is its Fourier transform in momentum space. $\epsilon^0_\bfk=\hbar^2\bfk^2/2m$ is the kinetic energy, and $\mu$ is the chemical potential. Eq. (\ref{E_FL}) can be also applied to systems of finite temperature, but in this paper we just consider zero temperature situation. In Fig. \ref{fermi_gas}(a), we show the calculated density distribution for different aspect ratios of the trap, and in (b) and (c) we show the obtained cloud's aspect ratio (normalized by the non-interacting aspect ratio). Here $R_{z}\equiv\left(\frac{1}{N} \int_{\bfr,\bfk}f_\bfk(\bfr)z^2\right)^{1/2}$ and $K_{z}\equiv\left(\frac{1}{N}\int_{\bfr,\bfk}f_\bfk(\bfr)
k_z^2\right)^{1/2}$ are the averaged radii in the spatial and momentum space (similar definitions for $R_x$ and $K_x$).  We find that our results
are consistent with Ref. [\oncite{HanPu}] for spatial aspect ratio, but disagree with them in the momentum aspect ratio. It is probably because their variational approach did not capture the full contribution of Hartree energy.

From Eq. (\ref{E_FL}), one can see that the susceptibility has two contributions: one is from the self-energy in Eq. (\ref{E_FL}), and the other is from the single particle energy via the deformation of Fermi surface. In low temperature limit, the latter contribution can be neglected, because it involves only an integral on the Fermi surface of the phase space. Besides, most effects of the self-energy in $f_\bfk(\bfr)$ can be absorbed by the shift of chemical potential due to the conservation of total number of particles. As a result, we can just keep the interaction effect to the leading order of $a_d(\theta)$, which is to replace $m|\langle\bfd\rangle|^2/\hbar$ in $V_d(\bfr)$ of Eq. (\ref{E_FL}). The first order transition does not exist here for the similar reason, since the high order behavior is enhanced more strongly in a bosonic gas. After re-scaling the length scale to $r_F\equiv\sqrt{2\varepsilon_F^{\rm ho}/m\omega_x^2}$ and the momentum scale to $k_F\equiv\sqrt{2m\varepsilon_F^{\rm ho}/\hbar^2}$ with $\varepsilon_F^{\rm ho}\equiv (6N)^{1/3}\hbar\bar{\omega}$ [\oncite{BEC_book}], we obtain (taking $\Omega\to 0$),
$E_{\rm orb}^{\rm FL}(\theta)N^{-1}=E^{\rm FF}_{\rm orb}N^{-1}
-48\sqrt{3N}\sin^2\theta (d_0^2/\bar{a}^3)C(\kappa)$,
where $E^{\rm FF}_{\rm orb}$ is the energy of free fermions.
$C(\kappa)\equiv\frac{4}{\kappa}\sqrt{\frac{\pi}{5}}
\int_{\tlbfr,\tlbfk}'\int_{\tlbfr',\tlbfk'}'\left[Y_{20}(\hat{\bfe}_r)|\bfr-\bfr'|^{-3}
+\frac{4\pi}{3}Y_{20}(\hat{\bfe}_k)\delta(\tlbfr-\tlbfr')\right]$ 
\begin{figure}
\includegraphics[width=8cm]{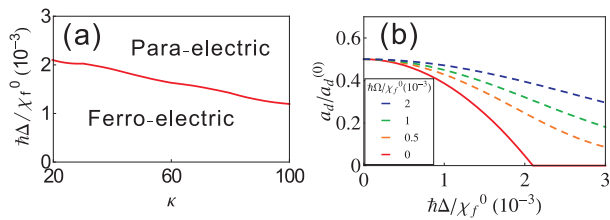}
\caption{Results of fermionic polar molecules: (a) Quantum phase diagram in terms of trap's aspect ratio, $\kappa$, and the detuning $\Delta$, in the zero field limit. (b) The calculated dipolar length, $a_d(\theta)$, as a function of $\Delta$. Results for different Rabi frequencies are shown together. 
}
\label{phase_fermion}
\end{figure}
is a dimensionless function of $\kappa$ with $Y_{lm}(\hat{\bfe}_{r,k})$ being the 
spherical harmonics. $\hat{\bfe}_{k}$ is the unit along
$\bfk-\bfk'$. $\int_{\bfr,\bfk}'$ stands for
the integral inside an elliptic sphere, $\tlx^2+\tly^2+\tlz^2/\kappa^2+\tlbfk^2=1$.
Note that $C(\kappa)$ plays a similar role as $A_2(\beta)$ in 
Eq. (\ref{E_BEC}). Therefore, expanding the total energy to the quadratic order of $\theta$, we can obtain the "susceptibility" per particle to be: 
$\chi_f\equiv\chi_f^0C(\kappa)$, where
$\chi_f^0\equiv 48(3N)^{1/2}d_0^2/\bar{a}^3$. The critical detuning for ferro-electricity is given by $\Delta_c=4\chi_f/\hbar$ as shown in Eq. (\ref{E_tot_b}).
For $^{40}$K$^{87}$Rb molecules with $N=10^4$, $\omega_\rho=2\pi\times 100$Hz, $d_0=0.5$ Debye and $\kappa=100$, we have $\chi_f^0=44.4$ kHz,
and $C(\kappa)=3\times 10^{-4}$ and hence $\Delta_c=4\chi_f/\hbar\sim 53$ Hz. This value is smaller than the corresponding value of bosonic molecules, because the condensate enhances the dipolar interaction effects. In Figs. \ref{phase_fermion}(a) we show the calculated quantum phase diagram, which is qualitatively similar to the bosonic case (see Fig. \ref{phase_boson}(d)). In Fig. \ref{phase_fermion}(b), we show the dipolar strength as a function of detuning. Results for different Rabi frequencies are also shown together.

Finally, we note that the ferro-electric order predicted here can be easily observed from the electric field distribution when the AC field is turned off adiabatically ($\alpha\to 0$). The resulting macroscopic polarization can also be manipulated by changing the confinement's aspect ratio, $s$-wave scattering length (bosonic case), or detuning frequency. Such macroscopically coherent rotational state can be neither reproduced in a single molecule picture nor in a uniform system. Finally, the "spin"-wave fluctuation as well as the texture in the rotational state shall also be an interesting subject for future study.  

In conclusion, we have shown that an interaction driven quantum
phase transition from a para-electric to a ferro-electric quantum
gas of polar molecules can be realized in the presence of a micro-wave field. We study the nature of such quantum phase transition, and then propose how to manipulate the molecular rotational states by the tuning of many-body parameters. 

The authors appreciate discussion with E. Demler, Y.-C. 
Chen, J. Bohn, S. Ronen, K.-K. Ni, H. Pu, and C. Sa de Melo. This work was supported by NSC (Taiwan).



\begin{thebibliography}{99}

\bibitem{JILA}
K.-K. Ni, {\it et al.}, Science {\bf 322}, 231 (2008);

\bibitem{others}
L.D. Carr and Jun Ye, New J. Phys. {\bf 11} 055009 (2009).

\bibitem{demille}
D. DeMille, Phys. Rev. Lett. {\bf 88}, 067901 (2002).

\bibitem{zoller}
H. P. Buchler, {\it et al.}, Phys. Rev. Lett. {\bf 98}, 060404 (2007);
A. Micheli, {\it et al.}, Phys. Rev. A {\bf 76}, 043604 (2007); G. Pupillo, A. Micheli, H.P. Buchler, P. Zoller, {\it Cold molecules: Creation and applications}, edited by R. V. Krems, B. Friedrich and W. C. Stwalley, and published by Taylor \& Francis. 

\bibitem{wang_dipolar_liquid_Santos}
D.-W. Wang, M. D. Lukin, and E. Demler, Phys. Rev. Lett. {\bf 97}, 180413 (2006); A. Arguelles and L. Santos, Phys. Rev. A {\bf 75}, 053613 (2007);
C. Kollath, J.S. Meyer, and T. Giamarchi, Phys. Rev. Lett. {\bf 100}, 130403 (2008); C.-K. Chan, {\it et al.}, arXiv:0906.4403.

\bibitem{ferro_Melo}
M. Iskin, and C.A.R. Sa de Melo, Phys. Rev. Lett. {\bf 99}, 110402 (2007).

\bibitem{eugene}
Maybe the only exceptions are the following two papers:
P. Rabl and P. Zoller, Phys. Rev. A, {\bf 76}, 042308 (2007), and
R. Barnett, {\it et al.}, Phys. Rev. Lett. {\bf 96}, 190401 (2006).

\bibitem{spinor}
S. Yi and H. Pu Phys. Rev. Lett. {\bf 97}, 020401 (2006), 
Y. Kawaguchi, H. Saito, and M. Ueda, Phys. Rev. Lett. {\bf 97}, 130404 (2006).

\bibitem{note_ferro}
The ferro-electricity proposed in Ref. [\oncite{ferro_Melo}] is based on a given dipole moment, which is not self-consistently determined by the many-body effects. Therefore it is {\it not} a spontaneous macroscopic order and disappears in the zero field limit.

\bibitem{multiferro}
W. Eerenstein, N.D. Mathur, and J.F. Scott, Nature {\bf 442}, 759 (2006);
S.-W. Cheong, and M.V. Mostovoy, Nature Mater. {\bf 6}, 13 (2007).

\bibitem{note}
Here we assume the AC field can be still described by a classical wave
in the limit of zero $\Omega$.


\bibitem{floquet}
See, for example, M. Frasca, Phys. Rev. A {\bf 60}, 573 (1999).

\bibitem{3d_dipolar_wang}
D.-W. Wang, New J. Phys. {\bf 10}, 053005 (2008).

\bibitem{Cr}
L. Santos, {\it et al.}, Phys. Rev. Lett. {\bf 85}, 1791 (2000);
For a review of dipolar condensate, see T. Lahaye, {\it et al.}, Rep. Prog. Phys. (to be published) 

\bibitem{BEC_book}
C.J. Pethick and H. Smith, {\it Bose-Einstein Condensation in Dilute gases}, Cambridge, new York (2002).

\bibitem{texture}
T.-L. Ho, Phys. Rev. Lett. {\bf 81}, 742 (1998);
M. Takahashi, {\it et al.}, Eur. Phys. J. B {\bf 68}, 391 (2009).

\bibitem{fetter}
A.L. Fetter and J.D. Walecka, {\it Quantum Theory of Many-Particle Systems}, Dover Publications (2003).

\bibitem{Bohn}
S. Ronen and J. Bohn, arXiv:0906.3753;
B.M. Fregoso, {\it et al.}, New J. Phys. {\bf 11}, 103003 (2009);
C.-K. Chan, {\it et. al.}, arXiv:0906.4403.

\bibitem{HanPu}
T. Miyakawa, T. Sogo, and H. Pu, Phys. Rev. A {\bf 77}, 061603(R) (2008);
T. Sogo, {\it et al.}, arXiv:0812.0948.

\end{thebibliography}
\end{document}